# On the action for a charged particle moving in a magnetic field

Manoj K. Harbola

Department of Physics

Indian Institute of Technology, Kanpur

Kanpur – 208016 (India)

**Abstract**

Application of the Bohr-Wilson-Sommerfeld quantization condition to a charged particle in a uniform magnetic field requires knowledge of the canonical momentum of such a particle, which in turn requires students to know about the vector potential. The canonical momentum in this situation is conventionally obtained by introducing an appropriate additional term involving the corresponding vector potential in the Lagrangian. In this work, we take a different approach to this problem by analyzing it with Maupertuis' principle of least action. We show that satisfying this principle for a charged particle moving in a uniform magnetic field requires that a term proportional to the flux passing through the area enclosed by the particle's trajectory be included in its action. This additional term accomplishes two tasks. First, it facilitates applying the Bohr-Wilson-Sommerfeld quantum condition for a charged particle moving in a uniform magnetic field without using the Lagrangian approach, and, secondly, it gives the appropriate canonical momentum and Lagrangian for a charged particle in a general electromagnetic field in a straightforward manner.

# I. Introduction

The principle of least action is fundamental across all of physics. One of the earliest examples of this principle is that of the principle of least distance, proposed by Hero of Alexandria. [1] More familiar to undergraduates will be Fermat's principle of least time, which states that a light ray traveling between two fixed points follows the path that takes the least time. [2-4] Students are also introduced to Maupertuis' principle of least action for a particle moving in a force field, particularly because it is needed to apply the Bohr-Wilson-Sommerfeld (BWS) quantum condition [5-9]. The action used in this principle is defined in Eqs. (1a) and (1b) below. The BWS quantum condition states that for a particle performing periodic motion, the corresponding action for one period of motion is equal to $nh$, where $n$ is an integer and $h$ is Planck's constant. In older quantum theory, the BWS condition is generally applied to obtain energies of a particle moving in a one-dimensional potential. [8] For example, application to a simple harmonic oscillator of angular frequency $\omega$ confirms Planck's proposal of the energy of the oscillator being $n\hbar\omega$, where $\hbar = h/2\pi$. Furthermore, it can be shown that Bohr's rule for quantization of angular momentum for the hydrogen atom also follows from this general condition. Notably, the BWS quantum condition paved the way for the development of quantum mechanics: its differential (difference) form was employed by Heisenberg to develop matrix mechanics. [10-12] A detailed discussion of the BWS quantum condition and its role in the development of matrix mechanics is given in January 2025 issue of this journal celebrating the International Year of Quantum Science and Technology. [13]

Although semiclassical in nature, results obtained from the application of the BWS quantum condition in various situations are helpful in understanding the quantized motion of a particle. However, if the results of calculations along this line are not "as expected", it may indicate something is amiss. An example, which is the subject of this paper, is such a calculation for a charged particle moving in a uniform magnetic field. This reveals that the expression for Maupertuis' action for a particle in a scalar potential must be modified. As discussed in what follows, while the expression for action is straightforward for a particle moving in a scalar potential, it becomes more involved for a charged particle in a magnetic field because the force on the particle depends on its velocity and cannot be expressed as the gradient of a potential. In the present work, the action for such a particle is calculated directly in a way that is different from and simpler than existing methods.[14, 15] This makes the application of BWS quantum condition to obtain the quantized orbits and energies of a charge in a magnetic field more straightforward and leads to results consistent with those expected from the classical motion of such a particle.

The outline of this paper is as follows. In section II we define Maupertuis' action for motion of a point particle in a scalar potential. In this section we also introduce two ways of obtaining the Lagrangian for a charged particle moving in a magnetic field. One is to include terms in the Lagrangian so that the corresponding Lagrange's equation matches the correct classical equation of motion. Using this Lagrangian, expressions for canonical momentum and the corresponding action for such a particle are derived. The second is to demand form invariance of the Lagrangian under certain transformations; this leads to the same Lagrangian as the first method. We then discuss the principle of least action in section III and introduce the motivation for the present work, which is developed in section IV. Readers who are not familiar with Lagrangian mechanics may wish to proceed directly to section III.

## II. Lagrangian, canonical momentum, and Maupertuis' action for a charged particle in an electromagnetic field

The Maupertuis' action for a particle is given by

$$S = \int \vec{p} \cdot d\vec{l}, \qquad (1a)$$

where $\vec{p}$ is its canonical momentum and $d\vec{l}$ is a line element along its trajectory. The canonical momentum is obtained from the partial derivative of the Lagrangian with respect to velocity. Thus, each component of canonical momentum is the partial derivative of the Lagrangian with respect to that component of the velocity. For a point particle of mass $m$ moving in a scalar field of potential energy $U(\vec{r})$ with velocity $\vec{v}$, the Lagrangian is

$$L_0 = \frac{m\vec{v}^2}{2} - U(\vec{r}) \quad ,$$

which gives $\vec{p} = m\vec{v}$ and

$$S = m \int \vec{v} \cdot d\vec{l}. \qquad (1b)$$

The formula for action in this case can also be written as

$$S = m \int |\vec{v}| |d\vec{l}| , \qquad (1c)$$

since $d\vec{l}$ is in the same direction as $\vec{v}$. On the other hand, for a charged particle in a magnetic field, $\vec{p}$ is different from $m\vec{v}$. This is because the Lagrangian $L(\vec{r}, \vec{v})$ must be such that the corresponding Lagrange's equation gives the correct equation of motion with the velocity-dependent Lorentz force.

The Lagrangian $L$ for a particle of charge $q$ in an electric field $\vec{E}$ and a magnetic field $\vec{B}$ is constructed by satisfying Lagrange's equation

$$\frac{d}{dt}\left(\frac{\partial L}{\partial \vec{v}}\right) - \left(\frac{\partial L}{\partial \vec{r}}\right) = 0,$$

where a partial derivative with respect to a vector quantity means the corresponding gradient, consistent with the equation of motion

$$m\frac{d\vec{v}}{dt} = -\vec{\nabla}U(\vec{r}) + q(\vec{E} + \vec{v} \times \vec{B})$$

This is done by working backwards from the equation of motion employing the electrostatic potential $V(\vec{r})$ and the vector potential $\vec{A}(\vec{r}, t)$ so that $\vec{E} = -\nabla V - \partial \vec{A}/\partial t$ and $\vec{B} = \vec{\nabla} \times \vec{A}$. [14, 16] Using the identities $d\vec{A}/dt = \partial \vec{A}/\partial t + (\vec{v}.\vec{\nabla})\vec{A}$ and $\partial(\vec{v}.\vec{A})/\partial \vec{r} = \vec{\nabla}(\vec{v}.\vec{A}) = \vec{v} \times (\vec{\nabla} \times \vec{A}) + (\vec{v}.\vec{\nabla})\vec{A}$ (it follows from the general identity since $\vec{r}$ and $\vec{v}$ are independent variables), this gives

$$L = L_0 - qV + q\vec{v} \cdot \vec{A} \qquad (2)$$

The resulting canonical momentum and action are now $\vec{p} = \partial L/\partial \vec{v} = m\vec{v} + q\vec{A}$ and $S = \int (m\vec{v} + q\vec{A}).d\vec{l}$, respectively.

Another approach taken to obtain $L$ is by invoking form invariance of the Lagrangian under certain transformations containing charge density and current. [15] For this, $U(\vec{r})$ in $L_0$ is considered to be of mechanical (non-electrostatic) origin. Then one transforms $L_0$ by adding the full time-derivative $q \, d\Lambda(\vec{r}, t)/dt = q \, \partial \Lambda/\partial t + q\vec{v} \cdot \vec{\nabla}\Lambda$ of a function $q\Lambda(\vec{r}, t)$ to it, which does not affect the equation of motion, and demands that for charged particles, the original Lagrangian have terms in it that give it the same form as the transformed Lagrangian $L' =$ $L_0 + q\,\partial\Lambda/\partial t + q\vec{v}\cdot\vec{\nabla}\Lambda$. This necessitates adding terms $q\vec{v}\cdot\vec{A}$ and $-qV$ to $L_0$, and leads to the same Lagrangian as in Eq. (2). A feature of this approach is that the terms involving the temporal and spatial derivatives of the function $\Lambda(\vec{r},t)$ correspond to the gauge transformations of the scalar and vector potentials, respectively.

So far, this discussion has been in terms of the Lagrangian, which is the central quantity of interest for further calculations of canonical momentum and action. However, it is clear that working out the correct Lagrangian requires many steps. This prompts the question of whether it can be done in a simpler way. The answer is "yes", and to do so we focus on action itself and show that for a charged particle in an electromagnetic field, we can calculate the Lagrangian from the trajectory in a uniform magnetic field by employing the principle of least action.

In what follows, we first obtain the action for a point charge moving in a uniform magnetic field and then use it to find the canonical momentum and the Lagrangian of the particle. It is then shown that these expressions are valid for the particle moving in a general electromagnetic field. The arguments are independent of the approaches taken in refs. [14] and [15] and therefore give a different way of obtaining the action, canonical momentum, and Lagrangian for a charge in an electromagnetic field. We effect our derivation by using Maupertuis' principle of stationary action and making it consistent with the trajectory for a charged particle in a uniform magnetic field; thus, our approach is akin to constructing the Lagrangian starting from the equation of motion. [14] We then rewrite this action in terms of the vector potential and identify the canonical momentum for the particle from this expression. As shown in Appendix A, the action so obtained also gives the correct equation of motion for a charged particle moving in a general electromagnetic field. Furthermore, it leads to the correct Lagrangian in a simple way.

In the following section, we state Maupertuis' principle of least action and demonstrate it through application to the example of a particle performing circular motion in a plane under the force $F(r) = -kr$, where $r$ is the distance from the origin. This sets the stage for discussing the action for a charge particle in a uniform magnetic field. For this, we consider the particle moving in a plane perpendicular to the magnetic field and show that consistency between the path of the particle and the action principle requires inclusion of the term $q\Phi$ in the corresponding action, where $\Phi$ is the flux passing through the area enclosed by the orbit of the particle.

### III. Principle of least action and action for a particle moving in a scalar potential

The principle of least action (PLA) can be stated as:

*If two points $P$ and $Q$ are taken on the trajectory of the particle and action is calculated between these two points for this trajectory and for a slightly different virtual trajectory close to the true one with its end points also at $P$ and $Q$ while keeping the energy of the particle fixed, then the first order difference between the actions for the two trajectories vanishes.*

The vanishing of the difference of the actions between the real and the virtual trajectories can be shown to lead to the equation of motion; this is done in Appendix A. Here we demonstrate the principle through an example chosen to facilitate later discussion of the action for a charged particle in a magnetic field.

Before proceeding further, we note that although the principle is known as the principle of least action, it is in reality a principle of stationary action, as that is what is implied by vanishing of the change in action between the true and the virtual trajectories. To see this, consider a particle moving in a field-free region surrounded by an elliptical boundary centered at the origin. On hitting the boundary, the particle bounces off it without losing any energy. If a particle starts from the origin and returns to it after getting reflected from the boundary, it can follow two possible paths, along either the semimajor or semi minor axis of the ellipse. Since the speed of the particle is constant, the path along the semimajor axis has maximum action, while that along the semi-minor axis has minimum action. Thus, the action for both the paths is stationary. Similarly, if the particle is to travel from one focus of the ellipse to the other, it can do so by reflecting from the wall. In this case, all lines passing through the two foci and meeting at the boundary form possible trajectories for the particle, and they will all have the same action since the sum of their lengths is constant; thus, the action along all such paths is stationary. The path of minimum action in this case is the straight line connecting the two foci.

To get to our example, consider a particle of mass $m$ undergoing circular motion in a plane under the force $F(r) = -kr$, where $r$ is the distance of the particle from the origin. This is shown in figure 1.

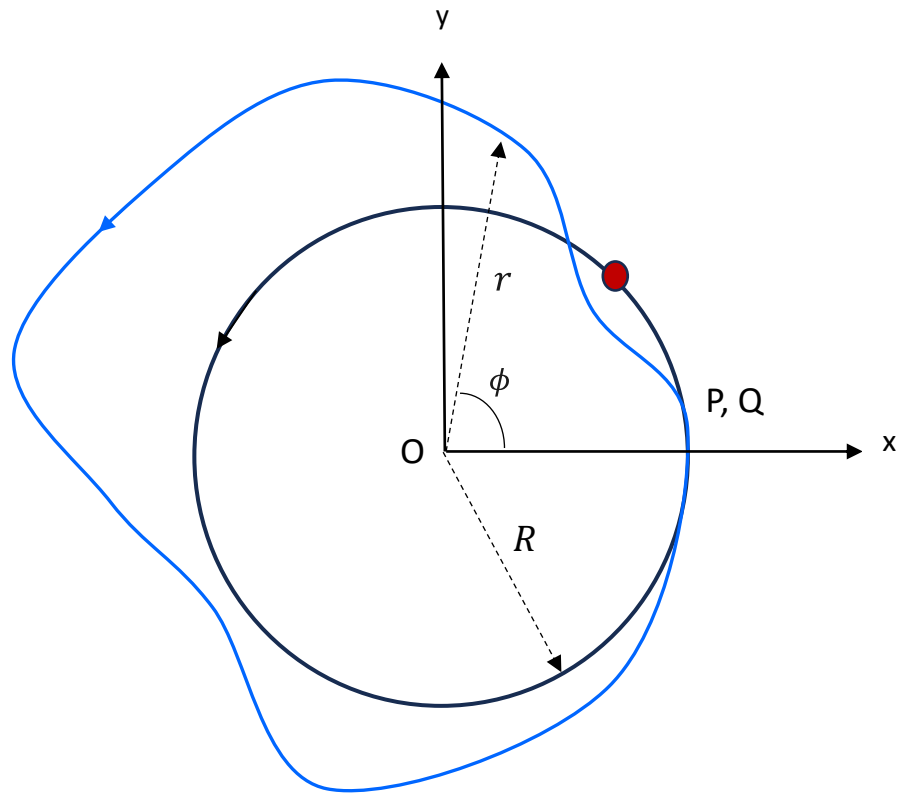

**Figure 1:** Circular trajectory of a particle, shown by red dot, under the influence of the force $F(r) = -kr$. The trajectory, shown in black, is centered at the origin and has radius $R$. Also shown in blue is a nearby virtual trajectory of arbitrary shape. Both trajectories start and end at the same points $P$ and $Q$, respectively.

The potential energy of the particle in this field is $kr^2/2$. Taking the speed of the particle to be $v_0$ and the radius of the circle to be $R$, the equation of motion is

$$\frac{mv_0^2}{R} = kR, \qquad (3a)$$

or

$$v_0 = \omega R \; ; \; \omega = \sqrt{\frac{k}{m}} \; . \qquad (3b)$$

The energy of the particle is

$$E = \frac{1}{2}mv_0^2 + \frac{1}{2}kR^2 = m\omega^2 R^2 \; . \qquad (4)$$

Note that the kinetic and potential energies are equal in this case as can be easily shown by the virial theorem.[19] The action integral for the motion over one period is now calculated, and it is shown that the action is stationary with respect to small deviations from the circular path.

Using Eq (1c), the action for one period of the motion is

$$S = 2\pi m v_0 R = 2\pi m \omega R^2 \; . \qquad (5)$$

Now consider a virtual path close to the circular path as shown by the blue curve in Figure 1. This has the same start and end points $P$ and $Q$ as the true trajectory. We work in polar coordinates with $\hat{r}$ and $\hat{\phi}$ being the corresponding unit vectors in the radial and the tangential directions, respectively. We can then write the equation for this path as

$$r(\phi) = R + af(\phi) \; , \qquad (6)$$

where $r(\phi)$ is the distance of a point on the virtual path at an angle $\phi$ from the x-axis with $f(\phi)$ being a well-behaved periodic function of $\phi$ with $f(0) = f(2\pi) = 0$. We take $a \ll R$ and consider terms up to the first order in $a$. The corresponding line element $\vec{dl}$ is

$$\vec{dl} = a\frac{df}{d\phi}d\phi \, \hat{r} + r\, d\phi \, \hat{\phi}. \qquad (7a)$$

Up to the first order in $a$, this gives

$$dl = \sqrt{a^2\left(\frac{df}{d\phi}\right)^2 + r^2}\, d\phi \cong r(\phi)d\phi. \qquad (7b)$$

Similarly, the velocity is

$$\vec{v}(\phi) = v_r \hat{r} + v_\phi \hat{\phi}$$

$$= a\frac{df}{d\phi}\dot{\phi}\,\hat{r} + r\dot{\phi}\,\hat{\phi}. \tag{8}$$

Here $v_r = a(df/d\phi)\dot{\phi}$ is the radial velocity and $v_\phi = r\dot{\phi}$ the tangential velocity on the virtual trajectory. Notice that $v_r$ is proportional to $a$, so up to the first order in $a$ we have

$$v(\phi) \cong v_\phi(\phi) \tag{9}$$

and

$$\vec{v}.\vec{dl} \cong rv_\phi d\phi$$

$$\cong rv(\phi)d\phi. \tag{10}$$

Keeping the energy of the particle fixed, its speed on the virtual trajectory up to first order in $a$ is

$$v(\phi) = v_0 - \omega a f(\phi). \tag{11}$$

The action integral is now

$$S = m\int_0^{2\pi} v(\phi)r(\phi)d\phi. \tag{12}$$

Up to the first order in $a$, Eqs. (6), (11), and (12) lead to the same action as that of Eq. (5) for the true trajectory: the first order change in action vanishes as required by the principle of least action. Thus, the trajectory of the particle obtained by applying the equation of motion, Eq. (3a), is consistent with the stationarity of action.

For periodic motion, an interesting relationship exists between the energy $E$, action per period $S$ and the frequency of motion $\omega/2\pi$. [20, 21] If $E$ is expressed as a function $E(S)$ with no other variables appearing in the function, then

$$\frac{dE}{dS} = \frac{\omega}{2\pi}. \tag{13}$$

In the present example, Eq. (4) and (5) give $E = \omega S/2\pi$, which satisfies this relationship.

The calculation of quantum-mechanical energy using the action obtained above in the BWS condition is discussed in Appendix B and yields the value $n\hbar\omega$. It is shown there how the correctness of the energy obtained is confirmed by applying Bohr's correspondence principle [21, 22], another guiding principle of the old quantum theory.

In the following section we repeat this exercise for a charged particle moving in a uniform magnetic field and show that action calculated by Eq. (1b) does not satisfy the principle of least action.

### IV. Action and Lagrangian for a charged particle moving in a magnetic field

**(a) Action**

Consider a particle of mass $m$ carrying positive charge $q$ moving with speed $v_0$ in the xy plane under the influence of a uniform magnetic field $B$ directed out of the page in the positive z-direction as shown in Figure 2. The direction of motion is clockwise. As is well-known, the solution of the equation of motion, which equates the centripetal force to the Lorentz force $qvB$ in this case, is that the particle moves in a circle of radius

$$R = \frac{mv_0}{qB} \tag{14}$$

with constant angular frequency

$$\omega = \frac{v_0}{R} = \frac{qB}{m} \quad . \tag{15}$$

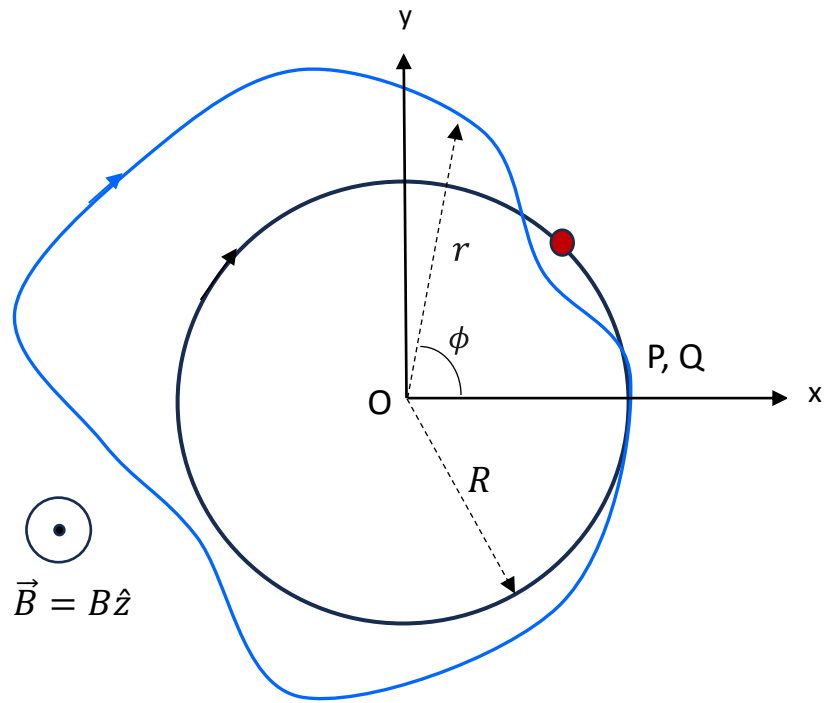

**Figure 2:** Circular path of a charged particle, shown by red dot, moving in a uniform magnetic field $\vec{B} = B\hat{z}$. The circle, shown in black, is centered at the origin and has radius $R$. Also shown in blue is a nearby virtual trajectory of arbitrary shape. Both the trajectories start and end at the same points $P$ and $Q$, respectively. Notice that the figure is the same as Figure 1 except that the particle now is moving in counterclockwise direction because of the direction of the Lorentz force that provides the centripetal acceleration.

The energy of the particle,

$$E = \frac{1}{2}mv_0^2 = \frac{1}{2}m\omega^2 R^2, \tag{16}$$

is purely its kinetic energy, and the Maupertuis' action per period for this motion calculated using Eq. (1b) or (1c) is

$$S_1 = 2\pi m v_0 R = 2\pi q B R^2 = 2q\Phi, \tag{17}$$

where $\Phi = \pi R^2 B$ is the magnetic flux passing through the area enclosed by the trajectory. But is this the correct value of the action? Classically, the energy of the particle in terms of action $S_1$ is $E = \omega S_1/4\pi$, which does not satisfy Eq. (13). This indicates that the action $S_1$ is not correct. Quantum-mechanically, it is shown in Appendix B that applying the BWS quantum condition to action $S_1$ gives the energy of the particle to be $n\hbar\omega/2$, which is inconsistent with the correspondence principle; this too indicates that the action given in Eq. (17) cannot be correct. [21, 22] As we will now show, this is because the value of action as given by Eq. (17) is not consistent with the principle of least action.

Consider a virtual trajectory shown by blue in Figure 2 near the circular trajectory of the particle. In the present case, if the energy of the particle is to be the same on this trajectory as that on the true one, the speed of the particle on the two trajectories will remain the same because the magnetic field does no work on the particle. Using first-order-in-$a$ approximations in Eqs. (9) and (10) gives the action calculated using Eq. (1c) to be

$$S_1 = 2\pi m v_0 R + m v_0 a \int_0^{2\pi} f(\phi) d\phi, \tag{18}$$

since $\vec{v} \cong -v_0 \hat{\phi}$. Also, with $\vec{dl}$ being parallel to $\vec{v}$, $\vec{dl} \cdot \hat{\phi} = -r(\phi) d\phi$.

Eq. (18) can be shown to follow in another way. Since the speed of the particle is $v_0$ over the entire trajectory, the action is the product of $mv_0$ and the length of the trajectory. By Eq. (7b), the length of the trajectory up to order $a$ is

$$\int dl = \int_0^{2\pi} r(\phi) d\phi \quad . \tag{19}$$

This leads to the same action as Eq. (18). However, it is clear from Eq. (18) that the action for the virtual trajectory is different from that for the true trajectory to within the first order in $a$. The difference between the action values for the two paths is

$$\Delta S_1 = m v_0 a \int_0^{2\pi} f(\phi) d\phi \qquad . \qquad (20)$$

which, unlike in the example of section III, does not vanish for arbitrary $f(\phi)$, disobeying the principle of action. This implies that with action defined as given by Eqs. (1b) and (1c), the principle of least action is not consistent with the trajectory obtained by solving the equation of motion. To make them consistent, the action for a charged particle in a magnetic field must be redefined.

To define the action for this problem, we add a term to the action given by Eq. (1b) that cancels the $\Delta S_1$ of Eq. (20) for the new trajectory. This term should be in terms of the magnetic field $B$ so that it does not appear in the expression for action when $B = 0$. To this end, we first replace $mv_0$ in Eq. (18) by $qBR$ and write

$$\Delta S_1 = qBRa \int_0^{2\pi} f(\phi) d\phi \qquad . \qquad (21)$$

In this expression, $Ra$ times the integral is the difference in the area between the two trajectories on keeping terms to first order in $a$ since

$$Area = \frac{1}{2} \int r(\phi) dl$$

$$\cong \frac{1}{2} \int_0^{2\pi} r^2(\phi) d\phi$$

$$= \pi R^2 + Ra \int_0^{2\pi} f(\phi) d\phi \qquad . \qquad (22)$$

This gives

$$\Delta S_1 = q\Delta\Phi \quad , \qquad . \qquad (23)$$

where $\Delta\Phi$ is the difference in the flux passing through the area enclosed by the true and virtual trajectories. It is evident that to make the change in the action vanish, the action for motion of a charged particle in a magnetic field must be defined as

$$S = m \int \vec{v} \cdot d\vec{l} - q\Phi \quad , \tag{24}$$

where $\Phi$ is the magnetic flux passing through the area enclosed by the particle's trajectory. On changing the path of the particle from the true to the virtual one, this gives the change in the action as

$$\Delta S = \Delta S_1 - q\Delta\Phi = 0 \quad , \tag{25}$$

which makes the principle of least action and the equation of motion consistent. Eq. (24) gives the correct action for the motion of charged particle in a uniform magnetic field to be

$$S = \pi m v_0 R = \pi m \omega R^2 \ . \tag{26}$$

With this value of action, the energy $E = \omega S/2\pi$, which is the same as in the example of section III and therefore satisfies Eq. (13).

We note that modifying the action for a charged particle in a magnetic field to make the action principle and the equation of motion compatible with each other is equivalent to including a term in the Lagrangian that makes the corresponding Lagrange's equation give the correct equation of motion.

As a simple example of the discussion above, take the virtual trajectory to be an ellipse of major axis $R + \delta_1$ and minor axis $R + \delta_2$ with both $\delta_1, \delta_2 \ll R$. For such an ellipse, the perimeter is $\pi(2R + \delta_1 + \delta_2)$ while its area, ignoring second-order terms in $\delta_1$ and $\delta_2$, is $\pi R^2 + \pi R(\delta_1 + \delta_2)$. This gives the result expressed by Eq. (23).

On going back to the problem of the quantized energy of a charged particle moving in a uniform field, the action $S$ in Eq. (26) when equated to $nh$ gives the energy of the particle to be $n\hbar\omega$, which is the correct energy; this is discussed in Appendix B.

A question arises why the two very different examples of circular motion considered in sections III and IV should have the same energy. This is addressed in Appendix B.

**(b) Canonical momentum and the Lagrangian**

We now define the canonical momentum for a charged particle in a magnetic field in terms of the magnetic vector potential $\vec{A}$. Recall that the magnetic field is obtained from the vector potential by the relation $\vec{B} = \vec{\nabla} \times \vec{A}$, and for the magnetic field $\vec{B} = B\hat{z}$ considered here, $\vec{A} = Br\hat{\phi}/2$. Canonical momentum is defined through Eq. (1a) and, therefore, we wish to put the action of Eq. (24) in the form of Eq. (1a). For this we use Stokes' theorem [23] and express the flux passing through the area enclosed by the particle's trajectory in terms of the vector potential as

$$\Phi = \int \vec{B} \cdot d\vec{S} = \oint \vec{A} \cdot \hat{\phi} R d\phi \qquad ,$$

which in turn gives

$$\Phi = -\int \vec{A} \cdot d\vec{l} \qquad . \qquad (27)$$

Here the negative sign arises because, being parallel to $\vec{v}$, the line element $d\vec{l} = -Rd\phi\hat{\phi}$. Then Eq. (24) can be written as

$$S = \int (m\vec{v} + q\vec{A}) \cdot d\vec{l} \qquad . \qquad (28)$$

Although we have reached the expression in Eq. (28) for a charge moving in a uniform magnetic field, it is shown in Appendix A that it is also the correct action for a charge in a general electromagnetic field as it leads to the correct equation of motion. Therefore, the canonical momentum of the particle is

$$\vec{p} = m\vec{v} + q\vec{A} \qquad . \qquad (29)$$

Furthermore, since the particle is moving without any constraints and its potential energy does not depend on its velocity, the Lagrangian for its motion is [24]

$$L = \vec{v} \cdot \vec{p} - E(\vec{r}, \vec{v})$$

where $E(\vec{r}, \vec{v})$ is the energy of the particle; see Appendix A for a derivation. With $E(\vec{r}, \vec{v}) = m\vec{v}^2/2 + U(\vec{r}) + qV(\vec{r})$, this gives the same Lagrangian as in Eq. (2). Thus, we have obtained the canonical momentum by satisfying the principle of least action.

## V. Concluding remarks

We have shown that for a charged particle moving in a uniform magnetic field, action has to be defined by including a term proportional to the magnetic flux passing through the area of its orbit in order to have consistency with the equation of motion. This makes the application of the Bohr-Wilson-Sommerfeld quantization condition for a charged particle in a magnetic field quite straightforward without introducing the concept of canonical momentum.  The action can then be used to define the canonical momentum and to obtain the Lagrangian for the motion.

**Conflict of interest:** The authors have no conflicts to disclose.

**Appendix A**

**Maupertuis' principle of least action, the equation of motion, and the Lagrangian**

In this Appendix we show that the principle of least action is equivalent to the equation of motion for a charged particle in an electromagnetic field.  The steps taken are similar to those given in ref. [17] but are generalized to include the vector potential.  The author is not aware of an explicit derivation of this kind in the case of Maupertuis' principle of least action for this situation.

Shown in figure A1 is the true trajectory and a nearby virtual trajectory of a particle. We wish to calculate the difference $\delta A$ in the action between the same end points $P$ and $Q$ on a section of these trajectories and show that taking this difference to be zero leads to the equation of motion.  From the definition of canonical momentum, the action for a particle in an electromagnetic field is

$$S = \int_P^Q (m\vec{v} + q\vec{A}) \cdot d\vec{l} \qquad . \qquad (A1)$$

The first-order change in action between the two trajectories is therefore

$$\delta S = m \int_P^Q \delta\vec{v} \cdot d\vec{l} + q \int_P^Q \delta\vec{A} \cdot d\vec{l} + m \int_P^Q \vec{v} \cdot \delta(d\vec{l}) + q \int_P^Q \vec{A} \cdot \delta(d\vec{l}) \qquad . \qquad (A2)$$

Here the difference denoted by $\delta X$ of a quantity $X$ gives its variation between the true and virtual trajectories and the differential $dX$ gives the infinitesimal change in $X$ along the trajectory. While the virtual difference $\delta X$ is at a given time, the differential $dX$ denotes the difference in values of $X$ at times $t + dt$ and $t$.  From the figure, it can be seen that

$$\delta(d\vec{l}) = d(\delta\vec{l}) \qquad , \qquad (A3)$$

because the differential change $d(\delta\vec{l})$ of the virtual displacement along the path is the difference $\delta(\vec{l} + d\vec{l}) - \delta(\vec{l}) = \delta(d\vec{l})$ in the virtual displacement at the two ends at $\vec{l}$ and $\vec{l} + d\vec{l}$ of the line element $d\vec{l}$. Using this we obtain

$$\int_P^Q \vec{v} \cdot \delta(d\vec{l}) = \int_P^Q \vec{v} \cdot d(\delta\vec{l}) = \int_P^Q d(\vec{v} \cdot \delta\vec{l}) - \int_P^Q d\vec{v} \cdot \delta\vec{l} \quad , \quad (A4)$$

and

$$\int_P^Q \vec{A} \cdot \delta(d\vec{l}) = \int_P^Q \vec{A} \cdot d(\delta\vec{l}) = \int_P^Q d(\vec{A} \cdot \delta\vec{l}) - \int_P^Q d\vec{A} \cdot \delta\vec{l} \quad , \quad (A5)$$

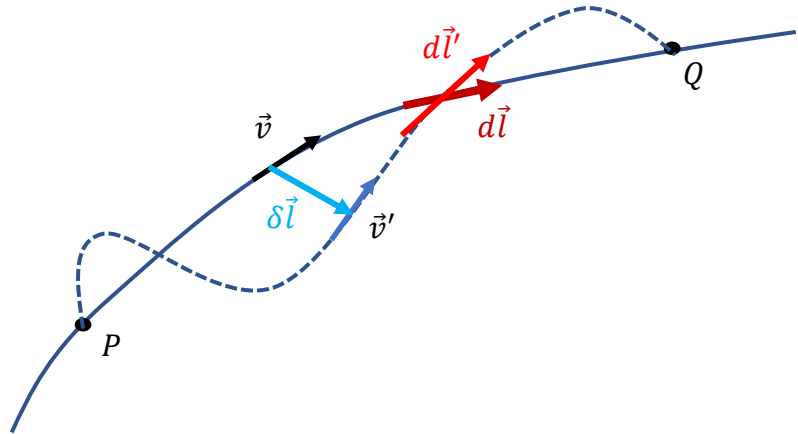

**Figure A1:** Shown as a solid line is a particle's true trajectory together with a virtual nearby trajectory (dashed line) obtained by virtually displacing points on the true trajectory. The virtual trajectory is made to pass through two points $P$ and $Q$ that lie on the true trajectory. Velocity $\vec{v}$ of the particle and an infinitesimal line element $d\vec{l}$ are shown on the true trajectory. The virtual displacement of a point along the virtual trajectory is depicted by $\delta\vec{l}$. The corresponding velocity on the virtual trajectory is $\vec{v}'$. The line element on the virtual trajectory corresponding to $d\vec{l}$ is $d\vec{l}'$.

Since $\delta\vec{l} = 0$ at the end points P and Q, the integrals $\int_P^Q d(\vec{v} \cdot \delta\vec{l})$ and $\int_P^Q d(\vec{A} \cdot \delta\vec{l})$ vanish and we get

$$\delta S = m \int_P^Q \delta\vec{v} \cdot d\vec{l} + q \int_P^Q \delta\vec{A} \cdot d\vec{l} - m \int_P^Q d\vec{v} \cdot \delta\vec{l} - q \int_P^Q d\vec{A} \cdot \delta\vec{l} \quad . \tag{A6}$$

We manipulate the first integral here using the fact that the particle's energy $E = m\vec{v}^2/2 + U(\vec{r}) + qV(\vec{r})$, where $U(\vec{r})$ is the mechanical scalar potential and $V(\vec{r})$ is the electrostatic potential. The energy on the two trajectories is to be kept the same, so $\delta E = 0$, which in turn implies

$$m\vec{v}.\delta\vec{v} = -\delta U(\vec{r}) - q\delta V(\vec{r}) \quad . \tag{A7}$$

We now make use of the increment of path length

$$d\vec{l} = \vec{v}dt \quad , \tag{A8}$$

where $dt$ is the time taken by the particle to traverse $d\vec{l}$, and also of the virtual changes of the potential energy and vector potential,

$$\delta U(\vec{r}) = \vec{\nabla}U(\vec{r}).\delta\vec{l} \quad \text{and} \quad \delta V(\vec{r}) = \vec{\nabla}V(\vec{r}).\delta\vec{l} \quad , \tag{A9}$$

$$\delta\vec{A} = \delta\vec{l} \cdot \vec{\nabla}(\vec{A}) = \hat{e}_j \delta l_i \partial_i A_j \quad , \tag{A10}$$

where $\hat{e}_j$ is a unit vector, with subscripts in all quantities denoting that direction and repeated indices imply summation over them. We then have

$$d\vec{A} = \left(\hat{e}_i v_j \partial_j A_i + \frac{\partial \vec{A}}{\partial t}\right) dt \quad , \tag{A11}$$

giving

$$\delta S = -\int_P^Q \left(\vec{\nabla}U(\vec{r}) + q\vec{\nabla}V(\vec{r})dt + md\vec{v} + q\frac{\partial \vec{A}}{\partial t}dt\right).\delta\vec{l}$$

$$+ q \int_P^Q (v_j \partial_i A_j - v_j \partial_j A_i)dt\delta l_i \quad . \tag{A12}$$

Now use $v_j \partial_i A_j - v_j \partial_j A_i = \epsilon_{ijk} v_j (\vec{\nabla} \times \vec{A})_k = (\vec{v} \times \vec{B})_i$ to write

$$\delta S = -\int_{P}^{Q}\left(\vec{\nabla}U(\vec{r}) + q\vec{\nabla}V(\vec{r})dt + md\vec{v} + q\frac{\partial \vec{A}}{\partial t}dt\right).\delta\vec{l} + q\int_{P}^{Q}(\vec{v}\times\vec{B}).\delta\vec{l}dt \quad . \quad (A13)$$

The principle of least action demands that $\delta S = 0$ for all virtual trajectories, which in turn implies $\delta S = 0$ for all possible $\delta\vec{l}$. This can happen only if the expression inside the brackets in the equation above vanishes. This leads to the equation of motion

$$m\frac{d\vec{v}}{dt} = -\vec{\nabla}U(\vec{r}) + q\vec{E} + q(\vec{v}\times\vec{B}) \quad , \quad (A14)$$

where $\vec{E} = -\vec{\nabla}V(\vec{r}) - \partial\vec{A}/\partial t$ is the electric field.

Now suppose that the energy is allowed to vary between different trajectories. In this case, Eq. (A7) appears as

$$m\vec{v}.\delta\vec{v} = \delta E - \delta U(\vec{r}) - q\delta V(\vec{r}) \quad . \quad (A15)$$

Therefore, we will get the correct equation of motion from $\delta S = 0$ if the action is taken to be

$$S = \int_{P}^{Q}(m\vec{v} + q\vec{A})\cdot d\vec{l} - \int_{P}^{Q}E dt \quad , \quad (A16)$$

with the time at the and points $P$ and $Q$ remaining fixed so that $\delta\int_{P}^{Q}Edt = \int_{P}^{Q}\delta Edt$. Using Eq. (A8), this gives

$$S = \int_{P}^{Q}\{(m\vec{v} + q\vec{A}).\vec{v} - E\} dt \quad . \quad (A17)$$

It follows from Hamilton's principle [18] that the Lagrangian is $L = (m\vec{v} + q\vec{A}).\vec{v} - E(\vec{r},\vec{v})$.

**Appendix B**

**Two principles of the old quantum theory: Bohr-Wilson-Sommerfeld condition and Bohr's correspondence principle, and their application to the examples in the paper**

The Bohr-Wilson-Sommerfeld (BWS) quantum condition states that for periodic motion, Maupertuis' action per period is an integer multiple of Planck's constant $h$. A good introductory discussion with application to a simple harmonic oscillator is given in ref. [8], and advanced-level applications are discussed in ref. [9].

Bohr's correspondence principle states that in the limit of large quantum numbers, results of quantum mechanics match with those of classical theory. In particular, properties such as the frequency, polarization, and intensity of light obtained in quantum mechanics should be the same as those calculated by applying classical mechanics and electromagnetic theory when quantum numbers become large. The matching of frequency follows from Eq. (13) of the text with differential $dE$ and $dS$ in energy and action replaced by their differences $\Delta E = E_n - E_{n-1}$ and $\Delta S = h$, respectively, between quantum levels $n$ and $n-1$ [21]. The principle is illustrated in ref. [22] with the example of frequency of light emitted by a hydrogen atom. For advanced application of the principle, the reader is referred to ref. [21].

We now apply these principles to obtain the energy levels of the examples considered in sections III and IV. In section III, we calculated the energy $E$ of a particle of mass $m$ moving in circular orbits in the potential $kr^2/2$ and showed that it can be written in terms of the action $S$ as $E(S) = \omega S/2\pi$, where $\omega = \surd(k/m)$ is the angular frequency of particle's revolution in the orbit. Equating the action to $nh$ according to the BWS condition gives the energy for the $n^{th}$ level

$$E_n = n\hbar\omega \qquad , \qquad (B1)$$

where $\hbar = h/2\pi$. The question is: Is this the correct energy? The answer is provided by the correspondence principle.[20, 21] The classical result for the angular frequency of radiation by a charge revolving with angular frequency $\omega$ is $\omega$ only; no other higher harmonics of this frequency exist. Thus, the minimum energy difference between two quantum energy levels should correspond to this frequency. The energies calculated above give this result and therefore must correspond to the quantum-mechanical energies.

Let us now turn to the case of a particle of mass $m$ and charge $q$ moving in a uniform magnetic field of magnitude $B$ with speed $v_0$. Classically, this particle too moves with angular frequency $\omega = qB/m$ in a circular orbit of radius $R = v_0/\omega$. Thus, classically the charged particle will emit radiation of the same frequency. The action $S_1$ of the particle per period calculated according to Eq. (1b) is related to the energy of the particle as $E(S_1) = \omega S_1/4\pi$. Equating $S_1$ to $nh$ gives the energy for the $n^{th}$ level as

$$E_n = n\left(\frac{\hbar\omega}{2}\right) \quad . \qquad (B2)$$

This gives the frequency of radiation when the particle jumps from a level to the adjacent level as $\omega/2$, which does not satisfy the correspondence principle. The energies given by Eq. (B2) thus cannot be correct. On the other hand, if the action $S$ is calculated according to Eq. (24), then the energy is given as $E(S) = \omega S/2\pi$. When $S$ is equated to $nh$, this results in energy $E_n = n\hbar\omega$, which gives the frequency of radiation to be $\omega$, consistent with the correspondence principle.

Finally, we address the question of why the energies in the two examples discussed above are the same. In the first example there is a potential energy but in the second the energy is purely kinetic. The answer lies in the relationship between the energy and the action. In the first example the total energy is two times the kinetic energy since the kinetic and potential energies are the same by the virial theorem [18]. This makes the expression for energy in terms of the angular frequency and the radius twice as large as in the second example, as is evident from Eqs. (4) and (16). However, as Eqs. (5) and (26) show, the action in the first example is also larger by the factor of two in comparison to the second example, making the expression for energy in terms of the action the same in both the cases. This, together with the quantized value of action being $nh$ renders the quantum-mechanical energies to be the same in the two cases.